\documentstyle[11pt,IAU207_pasp,twoside,psfig]{article}
\def\edcomment#1{\iffalse\marginpar{\raggedright\sl#1\/}\else\relax\fi}
\def\ltsima{$\scriptscriptstyle \; \buildrel < \over \sim \;$}
\reversemarginpar

\begin{document}
\title{Metal-Poor Globular Clusters of the Milky Way and Environs}
\author{James E. Hesser and Jon P. Fulbright}
\affil{Dominion Astrophysical Observatory, Herzberg Institute of Astrophysics,
National Research Council, 5071 W. Saanich Road, Victoria, BC V9E 2E7, Canada}
 
\medskip
\begin{flushright}
{\it ``He aqu\'\i$\;$ los astros p\'alidos todos llenos de enigma''} \\
Pablo Neruda, {\it El hondero entusiasta}\\
\end{flushright}

\begin{abstract}
We review properties of the most metal-deficient globular clusters and halo 
stars in the local universe as of March 2001, with goals of examining present 
evidence for the duration of the initial burst of massive cluster formation in 
the Milky Way, determining when that burst occurred, and elucidating what 
appears to be limiting our present understanding.  Such issues bear upon many
topics to arise later in the Symposium.
\end{abstract}
\section{Introduction}
The most metal-deficient stars of the local universe provide an unparallelled 
opportunity to study material representative of a much earlier epoch, perhaps 
predating the formation of the prominent components of the highest redshift 
galaxies being discovered today. While the massive globular star clusters (GCs) 
offer distinct, well-known advantages in such research, individual stellar 
chemistry is a powerful tool in the quest to understand early Galactic history. 
Unusually comprehensive recent review volumes are those of Mart\'\i nez Roger, 
P\'erez Fourn\'on \& S\'anchez (1999), Carney (2001) and Harris (2001). 

As is well known, the metallicity distribution of the  137 
Milky Way GCs with [Fe/H] determinations is well described  by two Gaussian 
curves centered on [Fe/H]$=$$-1.6$ and $-0.6$ (Harris 2001) on the popular 
Zinn \& West (1984) metallicity scale. Of particular interest in this context 
is that the most metal-deficient clusters have [Fe/H]$\sim$$-2.3,$ whereas the 
most metal-deficient halo stars extend to metallicity values more than an order 
of magnitude lower (c.f., Carney et al. 1996). While not entirely a proper 
comparison, the most metal-poor GCs and halo stars exhibit compositions well 
below those of damped Lyman-$\alpha$ forest systems at $z>3$ 
(c.f., Prochaska \& Wolfe 2000), yet ironically many of the extremely 
metal-dificient Galactic objects likely to have formed at high redshift can be 
seen with binoculars.

Other  properties of the GC system offer important clues to its origin. At 
Galactocentric radii (R$_{\rm gc}$) beyond 10~kpc, the average M$_{\rm v}$ 
of those clusters with blue HB morphology is $-7.3$, while clusters exhibiting 
red horizontal-branch (HB) morphology have a corresponding value of $-4.8$ 
(van den Bergh 1996). Half-light radii for Galactic GCs show a dependence 
on Galactocentric radii, with the farthest ones generally being larger in 
diameter.  While exhibiting similar trends, the clusters of the Fornax dSph or 
the LMC have a different distribution (van den Bergh 2000a).  
van den Bergh (2000a,b) interprets the present data as evidence that the 
Galaxy, Fornax and the LMC were distinct systems when they 
started to form their GCs: it appears that somehow the GCs ``knew'' they were 
going to end up in their respective parent galaxy.

By correlating the spatial distributions and kinematics of Galactic halo
tracers, Hartwick (2000) argues that the spatial distribution of Milky Way
companion galaxies define a real triaxial structure reflecting the initial
conditions of the Galaxy's formation epoch.  The outer halo GCs form a nearly
oblate, flattened system whose minor axis is highly inclined to the present
rotation axis, and thus do not seem part of the same triaxial system.  
 
\section{Length of Initial Burst} 
How long did the initial burst of cluster formation last for the Milky Way?  
The answer, from relative ages of the most metal-deficient halo stars, can 
help distinguish among conceivable formation scenarios (e.g., fast collapse, 
leisurely buildup through mergers, etc.). While relative measurements are 
usually more reliable to make and more straightforward to interpret, 
controversy about relative cluster ages persists. 

Morphological properties of HBs have been extensively 
explored for their potential as a powerful tool for sensitive age 
discrimination among clusters.  
From plots of metallicity against the ratio of the difference in the number of 
blue and red HB stars to the total HB population, compared with 
isochrones from synthetic HB models for GCs, Y.-W. Lee, Demarque, \& Zinn (1994,
{\it et seq.}; hereafter LDZ) found evidence for an age range of about 4 Gyrs, 
or some 25--35$\%$ of the age of a 12--15 Gyr old Galaxy. The age range so 
inferred is present at all metallicities, and an age-$R_{\rm gc}$ relation was 
suggested. The concepts which emerged from this important work have exerted 
wide influence in thinking about how long it took to form or assemble the 
Galactic halo.  Many accept that evidence from GCs strongly favors a relatively 
drawn out formation, which fits nicely with growing support for the importance 
of hierarchical mergers during the formation of large galaxies. However, this 
compelling interpretation  of HB morphologies has been the subject of 
debate over the past seven or so years. In part 
this arises because stellar evolutionary models indicate that HB morphology is 
sensitive to differences in parameters that are well below present 
observational uncertainties.  Specific counterexamples to the LDZ 
analysis noted by, e.g., Stetson, VandenBerg, \& Bolte  (1997), are 
disconcerting. What seems to be the current situation?

Harris et al. (1997) compared comparably deep (V,V$-$I)  color-magnitude 
diagrams (CMDs) for the massive GCs M92 and NGC$\thinspace2419$, which are 
among the most metal-deficient GCs in the Galaxy. They were unable to detect 
any age difference in excess of 1~Gyr. Importantly, M92 resides in the inner 
part of the Galactic halo, while NGC$\thinspace2419$'s 
R$_{\rm gc}$$\sim$$90$~kpc 
places it in the outermost halo (roughly twice as far as the LMC).  Moreover, 
its large core and tidal radii are a strong indication that it has existed 
in the outer halo throughout its life, for otherwise dynamical forces would 
have modified its appearance. (Note that Shetrone, C\^ot\'e, \& Sargent 
2001 have recently determined, for the first time,  a metallicity for 
NGC$\thinspace2419$ from high-dispersion 
spectra of giants which confirms the validity of pairing it with M92).
This single comparison among archtypical metal-deficient clusters 
from different regions of the halo already provides stringent constraints on 
the duration of the formation of the most massive clusters early in Galactic 
(pre)history. Grundahl's (2001) comparisons of deep CMDs for M92, 
NGC$\thinspace5053$ and NGC$\thinspace5466$ in the Str\"omgren photometric 
system similarly found no evidence of age differences in excess of $\sim$1~Gyr.

An extensive study by Rosenberg et al. (1999) applied both techniques of
relative age determinations, the color-difference (between the lower giant 
branch and the turnoff) and luminosity-difference (between the turnoff and 
the HB), to a homogeneous data set for 35 GCs. They found all the 
clusters with [Fe/H]$<1.2$ to be old and coeval. Some clusters in the range 
$-1.2 \leq$ [Fe/H] $\leq -0.9$ appear to be up to 25\% younger than the 
otherwise co-eval oldest clusters, while the most metal-rich clusters studied 
appear to be co-eval. They infer that the GC formation process started at the 
same time throughout the halo to a distance of $\sim$$20$~kpc from the Galactic 
center, with the small population of significantly younger halo clusters found 
at R$_{\rm gc}$$>$$8$ kpc likely the result of later mergers. They further note 
that their evidence for the co-eval metal-rich clusters being $\sim$17\% 
younger than other GCs (that is,  for a mild age-metallicity relation) is 
particularly model dependent. J.-W. Lee et al.  (2001) have extended the 
Rosenberg et al. analysis with three extremely metal-poor clusters located in 
the innermost Galactic halo, further strengthening the evidence against an age 
range or of a trend in ages with R$_{\rm gc}$ among the most metal-deficient 
GCs in the Galaxy.

On the question of the existence of an age-metallicity relation, VandenBerg 
(2000) analyzed in a systematic manner high-quality CMD data from the 
literature for 26 GCs. He infers that unless [O/Fe] or [$\alpha$/Fe] are up 
significantly at the lowest metallicities, then a mild age-metallicity relation 
results. At any given metallicity, however, he found the data do not permit
an age range $\geq$ 10-15\%.

The famous ``second-parameter'' GCs at intermediate metallicities and large 
R$_{\rm gc}$  figure prominently in efforts to understand Galactic halo 
formation. Their great distances, and low masses (hence sparse CMDs) have 
challenged observations from the ground. Stetson et al. (1999) 
reported relative age comparisons from HST data by both CMD techniques of three 
such clusters, Pal 3,4 and Eridanus, with the inner halo clusters M3 and M5, of 
similar metallicities. They found no evidence for an age spread among the three 
outer-halo clusters. At face value, their data suggest that those clusters 
could be younger by $\sim$$1.5$~Gyr than their comparison clusters; a 
caveat in their result is that errors of 0.2 dex (which cannot be ruled out 
with present data) in [Fe/H] or [$\alpha$/H] could reduce or eliminate the 
inferred small age difference between the two groups. Finally, two independent 
studies that compared CMDs for seven old, metal-poor GCs in the Large 
Magellanic Cloud with those of comparable clusters in the Galaxy failed to 
detect any age differences between them in excess of $\sim$1~Gyr or so (Olsen 
et al. 1998, Johnson et al. 1999); constraints on the age range from such 
studies are also potentially limited by the quality of available spectroscopic 
abundance determinations for the LMC clusters.

In summary, there is evidence that a fraction of the intermediate-metallicity 
GCs in the Galactic halo may be younger than the bulk of the system. However, 
in recent years evidence from careful differential CMD comparisons suggests 
any age dispersion  among the bulk of the Galactic GCs (or those in our 
distant halo) remains below the observational detection limit at a given 
metallicity, while the evidence for a mild age-metallicity relation remains a 
more debated question. It thus appears that the bulk of the Galactic GC system, 
and especially the most metal-poor ones, formed quickly during the earliest 
phases of Galactic history.

\section{When Did Formation Begin?}

We now turn to the more difficult question of absolute ages 
or, at what redshift did metal-poor star formation begin in the Galaxy? Recent 
work by  Bergbusch \& VandenBerg (2001) interpreting the CMD of M92 with their 
new isochrones favors an age of $\sim$15--16~Gyr.  This would be reduced by 
$\sim$$10$\% if He or heavy-element diffusion were incorporated into their 
models. Selection of the appropriate mixing-length for the comparison continues 
to require judgement. Three field halo subgiants they concurrently studied 
appeared also to have ages $\geq$$15$~Gyr. 

Grundahl et al. (2000)  found an M92 age of $\sim$15~Gyr using Str\"omgren 
color-color (c, c$-$y) diagrams, which have the powerful advantage of being 
distance independent. However, their interpretation requires accurate knowledge 
of the T$_{\rm eff}$ scale, which they (and Bergbusch \& VandenBerg 2001) argue 
is well-established. They also compared M92 Str\"omgren color-color data with 
those of the field subdwarfs having [Fe/H]$<$$-2$ and concluded that the 
subdwarfs are the same age, $\sim$$16$~Gyr, as M92 before taking diffusion into 
account. 

Carretta et al. (2000) studied nine clusters along with a large sample of 
subdwarfs with good parallax measurements, and concluded, after deciding that
distances from the subdwarf fitting method are $\sim$$1\sigma$ too long,
that the mean age of 
their sample was 12.9$\pm$2.9~Gyrs (95\% confidence range). Further analysis 
in terms of cosmological models led them to conclude that the initial epoch of 
GC formation occurred around z$\sim$3.

\begin{figure} 
\centerline{\vbox{\psfig{figure=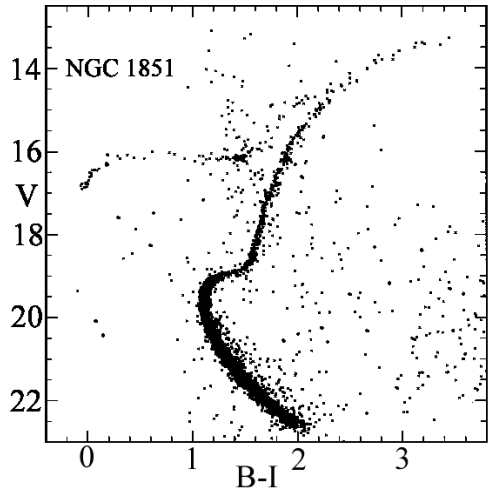,width=7cm,angle=0}}}
\caption{CMD by Stetson for NGC\thinspace1851 based upon 197 CCD images.} 
\end{figure} 
 
Until now, the quality--both internal and external--of photometry has been a 
limiting factor in the quest for reliable absolute ages for GCs. Fig. 1 is a 
remarkable example of work Peter Stetson has been doing in the past year or so 
combining data from many observers and telescopes. This particular CMD is based 
upon 197(!) CCD images from four telescopes obtained during 15 nights and eight 
observing runs. The data all came from the original observers. Enabling such 
research constitutes a powerful argument for cataloguing and archiving data 
from ground-based telescopes, as most images taken since the introduction of 
CCDs are tragically languishing on deteriorating tapes in observers' offices 
worldwide. With photometric zero points now known to better than 1\% on the new 
standard system (Stetson 2000), and with internal photometric 
scatter reduced to insignificance by combining many data, photometric quality 
will no longer limit the ability to determine accurate absolute and relative 
ages.  

\section{What Limits Our Present Understanding?} 

As photometric precision ceases to be a determining factor in the accuracy of 
age determinations, what factors remain important? Accurate knowledge of 
interstellar reddening for individual clusters and stars therein remains 
crucial, and nontrivial. Similarly, reducing uncertainties in distances from 
parallaxes remains a major challenge.  Uncertainties in chemical composition 
determinations are a fundamental yet extremely challenging subject ripe for 
renewed efforts, as we now discuss. For instance, there are still controversies 
about the basic scale of GC abundances (Zinn \& West 1984, Carretta \& Gratton 
1997; Rutledge, Hesser, \& Steton 1997). That highly experienced teams often 
find differences in overall abundances well beyond observational errors 
indicate 
that systematic errors still limit accuracy in high spectral resolution 
studies.  Finally, abundances for the most distant objects often are inferred
from broad-band photometry, or low-resolution spectra of $\alpha$-elements, 
like Ca, in the most luminous giants, rather than from high-dispersion spectra 
of less evolved stars.

\subsection {Abundances from giants and dwarfs} 

A question of considerable interest is whether inferences from more luminous 
cluster stars apply to the main-sequence and turnoff stars weighted so heavily 
in age determinations.  Until the advent of 8--10-m telescopes HB and red giant 
branch (RGB) stars have been the only ones for which abundances based upon 
higher spectral resolution were possible.

HB stars anchor distance moduli, underpin the LDZ 
interpretation of halo formation, and play many other important roles in our 
understanding of the galactic halo. an interesting example of modern abundance 
work is  Behr, Cohen, \& McCarthy's (2000) study of the abundances of HB stars 
in M15, with comparisons to M13 (Behr et al. 1999) and NGC\thinspace6752 
(Moehler et al. 1999). Among the slowly rotating hot stars 
two were found that exhibit the same extreme metal-deficiency that 
characterizes other M15 stars, but the other six stars studied show solar(!) 
[Fe/H] values due to the radiative levitation of metals predicted by Michaud, 
Vauclair, \& Vauclair (1983).  Many other interesting trends in detailed 
element abundances 
emerging from their study remind us how sensitively the HB reflects many 
parameters of stellar astrophysics.

The recognition of the phenomenon of abundance spreads among giant stars within 
an individual GC, but not among open cluster or field stars (e.g., Hesser, 
Hartwick, \& McClure 1976), sparked studies aimed at clarifying the percentage 
of the dispersion attributable to internal mixing during evolution up the giant 
branch, differences in initial abundances during the main-sequence phase, 
accretion, etc. Ivans et al. (1999) reported the range of individual chemical 
elements in the [Fe/H]$=$$-1.18$$\pm$$0.02$ cluster M4 determined from 
R=30,000 or 60,000 spectral data with S/N$>$100 for 36 giants. Most of the 
scatter seen for elements heavier than Al is observational, but they found 
evidence for a wide range of [O/Fe], [Na/Fe] and [Al/Fe] within this cluster.  
Since both [Fe/H] and [$\alpha$/Fe] play important roles in the age dating 
process, we need to understand how to characterize them for a given cluster 
when selecting appropriate isochrones.

Work with Keck and the VLT at higher resolution for fainter apparent magnitudes 
than heretofore possible using 4-m class telescopes (e.g., Cannon, et 
al. 1998) allowed Gratton et al. (2001) to compare abundances derived from
RGB and turnoff (TO) stars. Encouragingly, they found the same $<$[Fe/H]$>$ 
values from stars in these different evolutionary states in the clusters
NGC\thinspace6752 and 6397.  They also infer an O-Na anticorrelation among the 
main sequence stars, as previously seen among the giants in these clusters.

\subsection{[O/Fe], [Si/Fe] in Cluster and Field Stars}

As the third most abundant element, oxygen plays an important opacity role in 
stellar atmospheres.  Uncertainty in the [O/Fe] ratio is a factor in 
determining the absolute ages of GCs via theoretical isochrone fits.   
VandenBerg (1992) determined that a 0.25 dex change in [O/Fe] will change 
the resulting cluster age by 1 Gyr.  While modern abundance analyses can  
obtain internal relative abundance uncertainties of less than 0.25 dex, 
the systematic offsets between the different measurement methods 
of determining oxygen abundances can be much larger.   
 
\begin{figure}[t] 
\centerline{\vbox{\psfig{figure=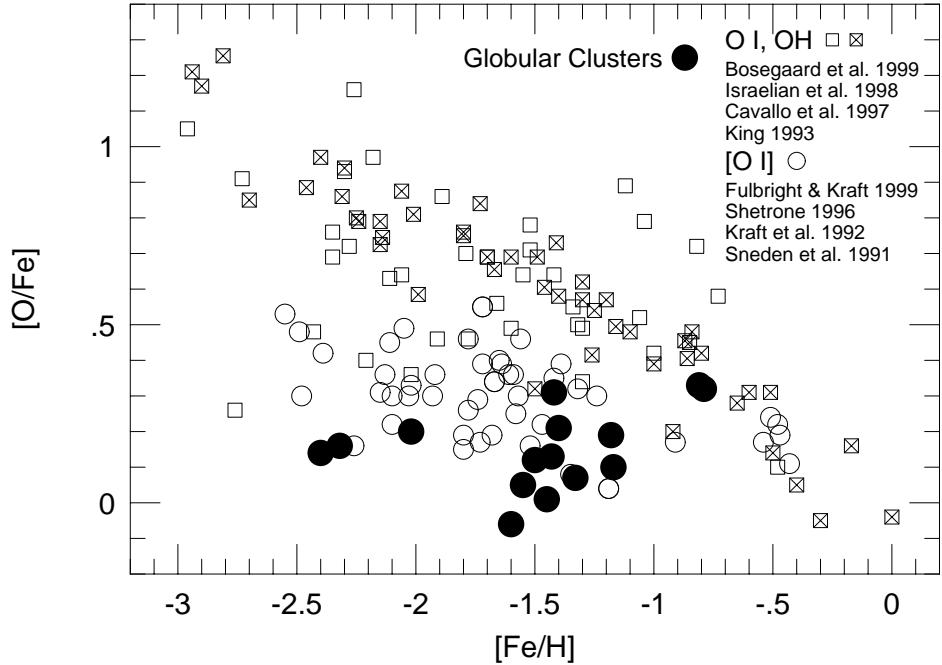,width=9cm,angle=0}}}
\caption{[O/Fe] ratios from several GC and field star studies.} 
\end{figure} 
 
Figure 2 plots [O/Fe] abundances determined by three methods:  the near-UV 
OH lines, the forbidden resonant [O I] lines in the optical, and the far 
red O~I permitted triplet.  Each method has strengths and drawbacks, as  
described in King (2000).  As can be seen in Figure 2 the oxygen  
abundance ratio measures determined from the [O I] lines are significantly 
lower than the determinations derived from the OH and permitted lines. 
For the most metal-poor GCs, the difference is about 0.5 dex, which translates
into a $\sim$2 Gyr systematic uncertainty in the ages of these GCs.
The GC oxygen abundances are determined from the forbidden line, which 
is relatively stronger in cool giants.  The permitted and OH lines are 
stronger in hotter stars, and the sample of very metal-poor stars with 
both forbidden- and permitted-line oxygen abundance analyses is small. 

Fulbright \& Kraft (1999), for example, found that the [O/Fe] ratio derived  
from [O I] line in two very metal-poor subgiants does not match the [O/Fe] 
ratio derived from the OH or permitted lines in the same stars. 
Recent attempts to resolve this discrepancy (King 2000, Israelian et al.  
2001) have involved a careful examination of the atmospheric 
parameters used  in the abundance analysis and the inclusion of 
non-LTE corrections.  Asplund \& Garc\'\i a P\'erez (2001) suggests that full 
3-D atmospheres may be necessary to model the oxygen lines correctly.
 
\subsection{[Si/Fe] between GCs} 

Silicon is the lightest $\alpha$-element that does not show large  
star-to-star abundance variations within individual metal-poor GCs.  However  
there does seem to be a cluster-to-cluster scatter in [Si/Fe]. 
As first pointed out by Kraft (2000), studies of M4  
(Ivans et al. 1999; 36 stars) and  
M5 (Sneden et al. 1992; 12 stars) found both to have similar values of 
[Fe/H] ($-1.18$ and $-1.17$, respectively), but M4 is twice as Si-rich as 
M5 ([Si/Fe] ratios of +0.55 and +0.20, respectively).  Both studies were 
conducted by the same group using similar techniques, and the star-to-star 
scatter within the individual clusters is smaller than the cluster-to-cluster 
difference. 
 
This is evidence that there was a significant chemical inhomogeneity between 
the material that formed these clusters.  Since the clusters are massive 
and on the metal-rich end of the halo metallicity distribution, it is unlikely 
that a random enrichment event would have had a sizable affect on the cluster's 
abundances.  The lack of [Si/Fe] variations within a cluster suggest that 
the cluster-forming material was well-mixed prior to formation, so whatever
led to the cluster-to-cluster variations in [Si/Fe] most likely took place
before the clusters formed.
 
\begin{figure}[t] 
\centerline{\vbox{\psfig{figure=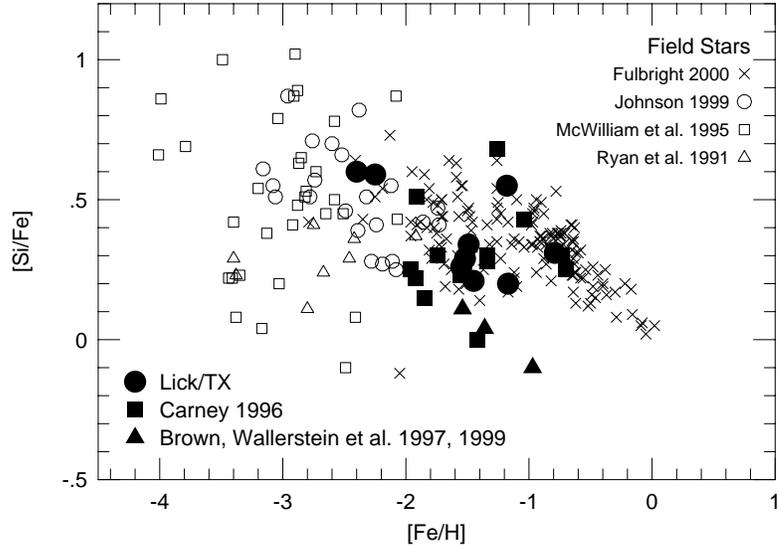,width=9cm,angle=0}}}
\caption{[Si/Fe] vs. [Fe/H] ratios from GC, dSph and field star studies.} 
\end{figure} 
 
In Figure 3 we plot the distribution of [Si/Fe] vs [Fe/H] for several 
globular clusters (mean values) and field stars. 
Aside from Ru 106 and Pal 12, there is general agreement in the range 
of [Si/Fe] values seen in three samples.  Two separate, parallel, tracks seem  
to be defined by M4 and the more metal-rich GCs and another by M5 and
the metal-poor GCs.
Whether this structure is truly there or is just a sampling effect is hard to 
determine from the data in hand.  Nissen \& Schuster (1997) found that  
for field stars with [Fe/H]$\sim$$-1$, there are halo stars that  
show lower [$\alpha$/Fe] enhancements than disk and other halo stars at 
a similar metallicity.  Fulbright (2001) shows evidence that for field 
halo stars with $-1$$>$[Fe/H]$>$$-2$, the stars having higher kinetic  
energy (e.g., on more extreme orbits) on average, show lower values of  
[X/Fe] for several elements, including [Si/Fe].   

Dinescu et al. (1999) computed orbital parameters for 38 GCs and found M4's 
apogalactic radius to be $\sim$6 kpc, while M5's is $\sim$40 kpc.   Thus 
connection between kinematics and observed abundances for halo objects extends 
to M4 and M5, as well.  The origin of such a relationship, however, is unclear.
The age information we have  
suggests that all the metal-poor GCs formed at about the same time, so both of 
these processes must have occurred simultaneously, although one or both of  
the clusters could have formed in a separate fragment that was accreted later. 
 
\subsection{Th and U: Independent paths to ages} 

An exciting new development in the study of GCs is the possibility that 
the long-lived radioisotopes of the r-process elements Th (14 Gyr half-life) 
and U (4.5 Gyr half-life) can be used to obtain ages for individual stars.   
Cowan et al. (1999) obtained a mean age of $15.6 \pm 4.6$ Gyr for two field 
stars based on Th line strengths.  Sneden et al. (2000) 
obtained a Th-based age for M15 of $14 \pm 3$ Gyr.  Johnson \& Bolte (2001) 
also measured Th-based ages for several field stars plus the M92 giant VII-18.  
For VII-18, Johnson \& Bolte obtain an age of $8.8$$\pm$$5.6$ Gyr.  

A major source of the large uncertainties is calculating the initial Th 
abundance for the stars.  In these studies the models for the initial r-process 
production were scaled to the observer ed stable rare earth elemental 
abundances (Eu, for example).   The mass differences between 
the rare earths and Th leaves room for sizable error.
 
Cayrel et al. (2001) successfully observed the 3895 \AA$\;$ line of U II 
in the metal-poor field giant CS31082-001 and derived an age of 
$12.5 \pm 3$ Gyr for it.  This star is special in that 
while the Fe and other light element abundances are about 1/800 of solar,  
the r-process abundances are about 1/9 of solar.  Having two radioactive 
species creates the advantageous situation where the observed U/Th ratio can 
be compared to a theoretical initial U/Th ratio without having to scale to 
other r-process elements.  Due to their proximity in atomic mass, the 
theoretical initial U/Th ratio should be more precise than scaling either to 
the lighter r-process elements (Goriely \& Clerbaux 1999).  
 
Besides the dependence on quality theoretical r-process calculations, the 
radioactive age-dating technique could be improved by work in two areas. 
First, our knowledge of the oscillator strengths of Th II and U II need  
improvement.  Cayrel et al. (2001) quote a 0.12 dex uncertainty in the U II 
log(gf) value they used.  That systematic error can change the resulting 
age by 2.6 Gyr.  It is hoped that the potential of this method to provide 
an independent age for the Galaxy will induce atomic physicists to improve 
the data. 
 
Second is the need for near UV/blue-sensitive very high resolution  
spectrographs on 10-m or larger telescopes.  While the present status 
of the field would not be possible without the excellent HIRES and UVES 
spectrographs, the near UV region of even these very metal-poor stars are 
still plagued with line blending at resolutions $\sim$50000--70000.  The  
situation would be vastly improved at resolutions $>$150000 with similar S/N 
levels. The reduction of blending problems and the increased contrast against 
the continuum would help make measurements of the Th and U lines possible 
in GC stars, which have not been found to be as enhanced in the r-process 
elements as CS31081-001. 

\section{Concluding Remarks} 
True to Neruda's poetry, the pale, metal-deficient objects throughout the 
Galaxy's extended environs remain enigmatic, but they continue  to yield 
important insights into how the Milky Way might have formed, and how that might 
in turn constrain understanding of galaxy formation in general. The salient 
points from  our attempt to review the current status are: 

1. The initial epoch of massive, metal-poor star cluster formation proceeded 
quickly ($<$1.5-2.0 Gyr); there are, however, younger clusters at 
intermediate-metallicities, which may have been accreted later. In particular, 
the most metal-deficient clusters (and field stars) show no discernible age 
dispersion throughout their 1-100 kpc extent in R$_{\rm gc}$. 

2. The most distant second-parameter GCs do not appear to be more than 
$\sim$$1.5$--2 Gyr younger than comparable metallicity, inner-halo clusters, 
but uncertain knowledge of their respective chemistry presents a caveat. 

3. Absolute ages of the most metal-deficient clusters remain controversial 
with recent determinations falling in the range 12 to 16 Gyrs with 
uncertainties of $\sim$2--3 Gyrs; recent consistency checks using abundances 
of radioactive elements (Th, U) are promising. 

4. Existence of a putative age-metallicity relation also remains controversial; 
it presently seems likely to be \ltsima 10\%. 

5. Photometric precision will soon not be a limiting observational factor on 
the accuracy of cluster ages; rather,  uncertainties in reddenings, distances 
and chemical composition will dominate. The time is ripe for major efforts on 
abundance determinations with 8-m and larger telescopes, but the most 
interesting problems may require telescopes with 20-30-m aperture. 

6. There is growing evidence that dispersion of light elements within 
individual GCs reflect the epoch of star formation, rather than of mixing 
during stellar evolution; such differences combine with uncertain [O/Fe] 
determinations to complicate accurate age determinations. It is encouraging 
that the average compositions determined from evolved and turnoff region stars 
seem to be the same within the limits of the data quality in the few clusters 
yet examined. 

7. Cluster-to-cluster differences for ratios of light elements imply 
substantially different enrichment histories.  

8. Field halo stars exhibit wide ranges of [X/Fe]; such differences provide 
detailed insight into early star formation. 

9. Over $\sim5$ orders of magnitude in total mass (from GCs to dSphs to the 
LMC), the most metal-poor Local Group constituents appear to have formed at the 
same epoch, yet to also have been aware of their ultimate association with a 
articular parent. 
 
We are grateful to Frank Grundahl, Jennifer Johnson, J.-W. Lee and  Peter 
Stetson for sharing material that we used when preparing this paper, and to 
Eva Grebel and 
Doug Geisler for organizing an outstanding conference in a beautiful location.

\section*{Discussion}

\noindent {\it D'Antona:\, } [COMMENT MADE AFTER DEMARQUE'S REMARKS] \\
Recently, Gratton et al. have found that the iron 
and oxygen abundances in NGC 6397 turnoff stars are equal to subgiant 
branch abundances, although models including diffusion predict differences,
also by a factor of two, for this metal-poor cluster.  Of course, this does
not give information on helium diffusion, which affects the age determinations,
but is an indicator that the models are not adequate for the diffusion of
metals. \\
 
\end{document}